\documentclass[rnote]{aa}  
\usepackage{graphicx}
\usepackage{txfonts}
\usepackage{fixltx2e}
\usepackage{lscape}
\begin{document} 
   \title{The far-infrared emission of the radio-loud quasar 3C~318}
   \author{P.~Podigachoski     
      \and P.~D.~Barthel
      \and R.~F.~Peletier
      \and S.~Steendam
          }

   \institute{Kapteyn Astronomical Institute, University of Groningen, 9747 AD Groningen, The Netherlands\\ \email{podigachoski@astro.rug.nl}\label{inst1}
              }
   \date{Received; accepted}
   
   \abstract{3C~318, a radio-loud quasar at $z$=1.574, is a subgalactic-sized radio 
             source, and a good test-bed for the interplay between black hole and 
             galaxy growth in the high-$z$ Universe. Based on its \textit{IRAS}, 
             \textit{ISO}, and SCUBA detections, it has long been considered as one 
             of the most intrinsically luminous (L$_{\mathrm{IR}}$ > 10$^{13}$ 
             L$_{\odot}$) infrared sources in the Universe. Recent far-infrared data 
             from the \textit{Herschel Space Observatory} reveal that most of the 
             flux associated with 3C~318 measured with earlier instruments in 
             fact comes from a bright nearby source. Optical imaging and spectroscopy 
             show that this infrared-bright source is a strongly star-forming pair of 
             interacting galaxies at $z$=0.35. Adding existing \textit{Spitzer} and 
             SDSS photometry, we perform a spectral energy distribution analysis of 
             the pair, and find that it has a combined infrared luminosity of 
             L$_{\mathrm{IR}}$ = 1.5 $\times$ 10$^{12}$ L$_{\odot}$, comparable to 
             other intermediate-redshift ultra-luminous infrared galaxies studied with 
             \textit{Herschel}. Isolating the emission from 3C~318's host, we robustly 
             constrain the level of star formation to a value a factor of three lower 
             than that published earlier, which is more in line with the star formation 
             activity found in other \textit{Herschel}-detected 3CR objects at similar redshift.                   
             }
   \maketitle
\section{Introduction}
\label{section:Introduction}
Powerful, high-redshift, radio-loud active galactic nuclei (AGN) provide a unique 
opportunity to study the interplay between the growth of the black hole and that of 
the host galaxy when both processes went through their peak activity. 

The Revised Third Cambridge Catalogue of radio sources \citep[hereafter 3CR;][]{Bennett62,Spinrad85} 
contains some of the most luminous radio galaxies (RGs) and 
quasars (QSRs) in the high-$z$ Universe, which have been studied with virtually all 
space telescopes \citep[e.g.][]{Best98,Haas08,Leipski10,Wilkes13,Chiaberge15}.
\citet[][hereafter P15]{Podigachoski15} recently performed a comprehensive study of this 
sample using far-infrared (FIR) data from the \textit{Herschel Space Observatory} 
\citep{Pilbratt10}, covering the last remaining spectral window in the study of these 
landmark objects. In that work, we found evidence for strong star formation (SF) 
activity in about 40\% of 3CR hosts, at the level of several hundred solar masses per 
year, comparable to the SF activity in equally massive, non-AGN hosts at similar redshift.    

3C~318 \citep{Spinrad&Smith76} is a QSR at a spectroscopically measured redshift of 
$z$=1.574 \citep{Willott00}. It is a compact steep-spectrum (CSS), presumably young 
radio source, with a projected radio extent of $\sim$7~kpc \citep{Mantovani10}. Adopting 
a typical speed of radio-jet expansion of about 10\% of the speed of light suggests that 
the jets were triggered $\sim$~0.1~Myr ago. 3C~318 is one of the highest redshift sources 
detected with the \textit{IRAS} telescope \citep{Hes95}, and is also detected with SCUBA 
at 850~$\mu$m \citep{Willott02}. These two strong detections suggested that 
3C~318 is unusually bright in the infrared. Given the steep-spectrum nature of the radio 
source, it is clear that the dominant emission in the infrared is thermal dust emission 
in its host galaxy. Taking into account the AGN contribution to the 850~$\mu$m flux, 
estimated from the synchrotron spectral slope, \citet{Willott07} reported a hyper-luminous star 
formation rate (SFR) of 1700 M$_{\odot}$ yr$^{-1}$. \citet{Heywood13} targeted the 
ground-state CO line in 3C~318, and found that the molecular gas with an estimated mass of 
$M_{H_2}$ $\sim$ 3.7 ($\pm$0.4) $\times$ 10$^{10}$ ($\alpha_{CO}$/0.8) M$_{\odot}$ is 
spatially offset from the position of the QSR. These authors argued that 3C~318 is either 
undergoing a major merger (whereby the starburst is taking place in a nearby merging 
galaxy), or is a highly-disturbed system. Such an amount of molecular gas combined with 
the estimated SFR result in a gas-depletion time scale of about 20 Myr in this system. 

As already noted by \citet{Spinrad&Smith76}, a faint pair of interacting galaxies is 
visible in their optical image\footnote{Note that the scale provided with their optical 
image is incorrect, and that their redshift of 3C~318 was later updated to be $z=1.574$ 
\citep{Willott00}.} about 20$\arcsec$ west of 3C~318. The \textit{r}-band SDSS-field of 
3C~318 is shown in Fig.~\ref{figure:stamps}, with the position of 3C~318 and the galaxy 
pair (East and West) marked. The pair emits prominently in the \textit{Herschel} maps 
published by P15. Given the poor angular resolution of \textit{IRAS}, it is clear that 
the pair contributes significantly to the measured \textit{IRAS} flux. Here, we assess 
the nature of the interacting galaxies by spectroscopically measuring their redshifts, 
and analysing their spectral energy distributions (SEDs). Using this newly acquired 
information, we update the SED of 3C~318's host, and modify several physical properties 
related to this peculiar object.    
\section{Existing and new observations of 3C~318 and the nearby galaxy pair}
\label{section:Data}
\begin{figure*}
\centering
\includegraphics[width=13cm]{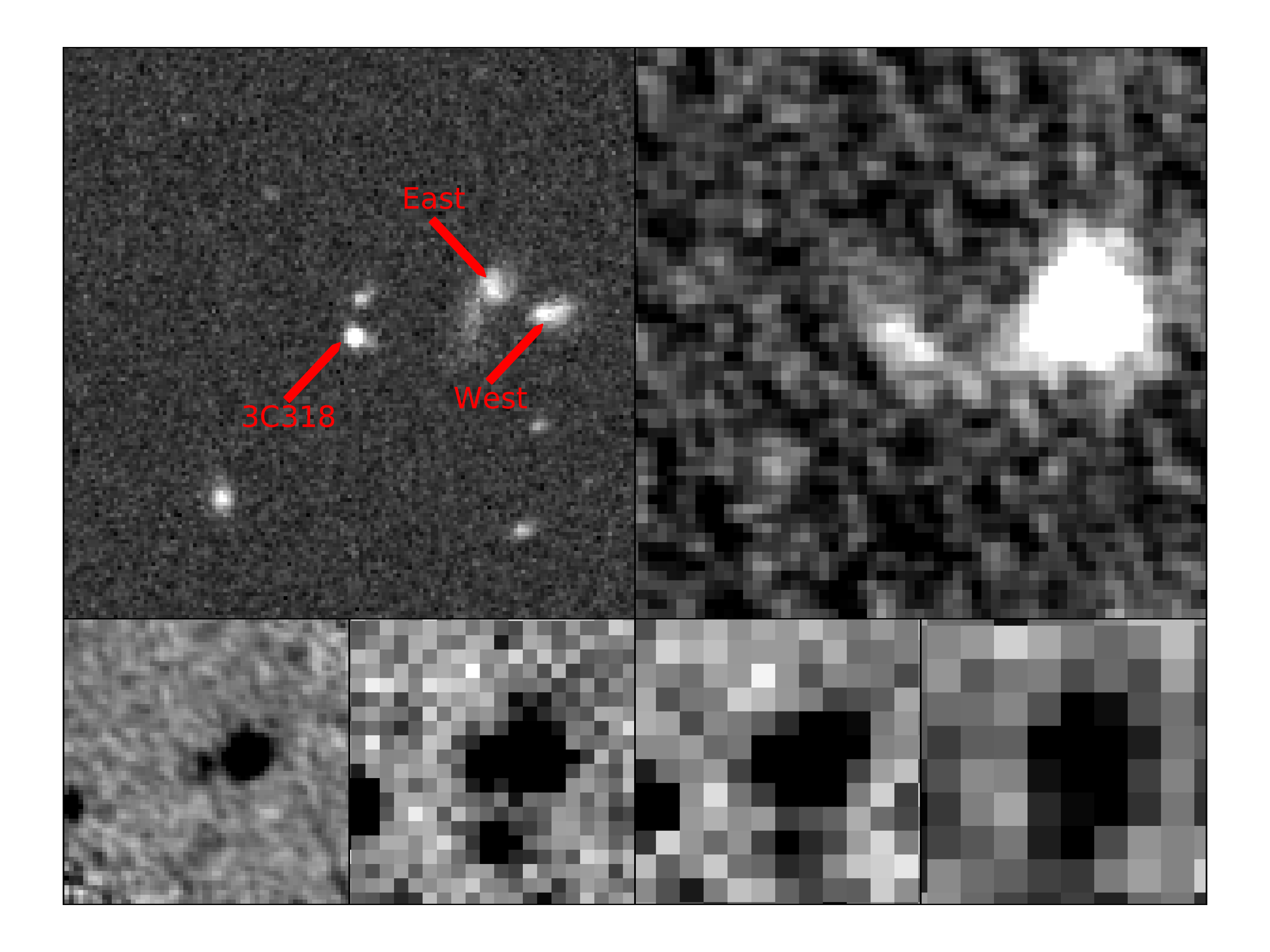}
\caption{3C~318 and the nearby pair of interacting galaxies. Top left: SDSS \textit{r}-band 
         image of 3C~318's field, showing 3C~318 and the East and West galaxies of the nearby 
         pair. Top right: PACS 70~$\mu$m image showing 3C~318's host and the infrared-bright 
         nearby pair. Each field is 1x1 arcmin in size and centered on 3C~318. Note that both 
         the optical and the infrared images indicate the presence of additional objects close 
         (in projection) to 3C~318. Lower panels: from left to right, \textit{Herschel} maps in 
         PACS 160~$\mu$m, SPIRE 250~$\mu$m, 350~$\mu$m, and 500~$\mu$m, respectively. These maps 
         are 2x2 arcmin in size and centered on 3C~318.
         }
\label{figure:stamps}
\end{figure*}
3C~318 is one of the strongest submillimetre (SCUBA) sources 
(F$_{850 \mu \mathrm{m}}$ = 7.8 $\pm$ 1.0 mJy) detected by \citet{Willott02}. Given that 
it is also detected with MAMBO \citep[F$_{1.25 \mathrm{mm}}$ = 5.6 $\pm$ 1.4 mJy;][]{Haas06}, 
it is clear that part of the SCUBA flux is contributed by non-thermal (synchrotron) emission 
from 3C~318's radio core. Extrapolating from core radio data, \citet{Haas06} estimated this 
contribution to be $\sim$~3.5 mJy, which means that about 4.3 mJy are due to thermal radiation 
from dust in 3C~318's host.  

3C~318 and its surroundings were observed with the \textit{Spitzer Space Telescope} 
\citep{Werner04} with all three imaging instruments: IRAC at 3.6, 4.5, 5.8, and 8.0~$\mu$m, 
IRS-16 peak-up array at 16~$\mu$m, and MIPS at 24~$\mu$m. Details of the data reduction and 
photometry were published by \citet{Haas08}. The photometry is presented in Table~\ref{table:photometry}.

The \textit{Herschel} observations of the field around 3C~318 were obtained as part of 
our Guaranteed Time project: The \textit{Herschel} Legacy of distant radio-loud AGN 
(PI: Barthel). Five-band photometric measurements were carried out using both imaging 
instruments, PACS \citep{Poglitsch10} at 70 and 160~$\mu$m, and SPIRE \citep{Griffin10} 
at 250, 350, and 500~$\mu$m. Data reduction was performed as described by P15. In 
Fig~\ref{figure:stamps}, 3C~318 can be clearly seen in the final PACS and SPIRE 250~$\mu$m maps, 
but the interacting galaxies are unresolved, and deblending of the pair is not possible. 
The galaxy pair and 3C~318 fall within the same beam in the 350 and 500~$\mu$m SPIRE bands. 
While 3C~318 can still have significant emission in the SPIRE 350 and 500~$\mu$m bands, it is 
evident that most of the FIR emission comes from the galaxy pair. The 3C~318 flux densities in 
the PACS bands were obtained as outlined in P15, using apertures of 4$\arcsec$ and 6$\arcsec$ 
radius, for PACS 70~$\mu$m and PACS 160~$\mu$m, respectively. Integrated PACS flux densities 
for the galaxy pair were measured using a 6$\arcsec$ and 12$\arcsec$ aperture in PACS 70~$\mu$m 
and 160~$\mu$m, respectively. SPIRE photometry followed the procedure adopted by P15. Given the 
steep radio spectrum of 3C~318, no synchrotron contribution is expected in the SPIRE bands. 
All flux densities measured from the \textit{Herschel} bands are listed in Table~\ref{table:photometry}. 

To determine the redshift of the pair of galaxies, and hence quantify the contribution of 
3C~318 to the integrated flux density in the 350 and 500~$\mu$m SPIRE bands, we obtained 
optical spectra for each of the two galaxies with the ACAM instrument on the William 
Herschel Telescope (WHT) on February 7, 2014. The long-slit spectra were reduced using 
standard packages within IRAF. Identifying emission lines from [\ion{O}{II}], \ion{H}{$\beta$}, 
[\ion{O}{III}], and \ion{H}{$\alpha$} in the noisy spectra, we measure the redshift for each 
galaxy to be $z$=0.35$\pm$0.01, with a relative velocity difference of $\sim$~300~km~$s^{-1}$. 
Additional optical data (\textit{ugriz} bands) for the pair of interacting galaxies were 
obtained from SDSS\footnote{\url{http://www.sdss.org/dr12/}}. The flux densities in the optical 
bands are also listed in Table~\ref{table:photometry}.
\begin{table*}
\renewcommand{\arraystretch}{1.2}
\caption{Photometry of 3C~318 and of the individual galaxies constituting the interacting pair 
         close (in projection) to 3C~318.}   
\label{table:photometry}      
\centering                      
\begin{tabular}{cccccc}      
\hline\hline
  Band                      & 3C~318         & East galaxy    & West galaxy    & Galaxy pair     & 3C~318 and pair \\    
\hline\hline          
  SDSS \textit{u} [$\mu$Jy] & ...            & 3 $\pm$ 2      & 6 $\pm$ 2      & ...             & ... \\              
  SDSS \textit{g} [$\mu$Jy] & ...            & 26 $\pm$ 1     & 15 $\pm$ 7     & ...             & ... \\               
  SDSS \textit{r} [$\mu$Jy] & ...            & 72 $\pm$ 2     & 48 $\pm$ 1     & ...             & ... \\               
  SDSS \textit{i} [$\mu$Jy] & ...            & 108 $\pm$ 2    & 77 $\pm$ 1     & ...             & ... \\               
  SDSS \textit{z} [$\mu$Jy] & ...            & 141 $\pm$ 10   & 112 $\pm$ 7    & ...             & ... \\               
  IRAC 3.6 [$\mu$Jy]        & 343 $\pm$ 51   & 247 $\pm$ 24   & 289 $\pm$ 28   & ...             & ... \\     
  IRAC 4.5 [$\mu$Jy]        & 427 $\pm$ 64   & 238 $\pm$ 23   & 293 $\pm$ 29   & ...             & ... \\
  IRAC 5.8 [$\mu$Jy]        & 571 $\pm$ 86   & 212 $\pm$ 21   & 315 $\pm$ 31   & ...             & ... \\
  IRAC 8.0 [$\mu$Jy]        & 806 $\pm$ 121  & 540 $\pm$ 54   & 933 $\pm$ 93   & ...             & ... \\
  IRS  16 [$\mu$Jy]         & 1960 $\pm$ 294 & 540 $\pm$ 54   & 1500 $\pm$ 150 & ...             & ... \\ 
  MIPS 24 [$\mu$Jy]         & 3400 $\pm$ 510 & 1140 $\pm$ 114 & 3000 $\pm$ 300 & ...             & ... \\ 
  PACS 70 [mJy]             & 17.2 $\pm$ 1.8 & \textit{58}    & \textit{163}   & 221.2 $\pm$ 3.8 & ... \\ 
  PACS 160 [mJy]            & 38.3 $\pm$ 4.2 & \textit{73}    & \textit{205}   & 277.5 $\pm$ 6.1 & ... \\ 
  SPIRE 250 [mJy]           & 35.9 $\pm$ 6.1 & \textit{43}    & \textit{120}   & 163.4 $\pm$ 6.1 & ... \\ 
  SPIRE 350 [mJy]           & \textit{11}    & ...            & ...            & ...             & 91.3 $\pm$ 6.2 \\ 
  SPIRE 500 [mJy]           & \textit{16}    & ...            & ...            & ...             & 46.6 $\pm$ 6.3 \\ 
  SCUBA 850 [mJy]           & \textit{4.3}   & ...            & ...            & ...             & ... \\ 
\hline           
\end{tabular}
\tablefoot{The East (15:20:04.45, +20:16:10.55) and West (15:20:04.05, +20:16:07.91) 
           galaxies are unresolved in the PACS 70~$\mu$m, PACS 160~$\mu$m, and SPIRE 250~$\mu$m bands. 
           The \textit{Herschel} PACS and SPIRE OBSIDs are 1342223844/1342223845 and 1342204107, 
           respectively. Both galaxies and 3C~318 are within the same beam in the SPIRE 350~$\mu$m and 
           SPIRE 500~$\mu$m bands. Estimated photometry (as explained in the text) is given in italics.
          }
\end{table*}
\section{Results and discussion}
\label{section:Results}
\begin{figure}
\centering
\includegraphics[width=\hsize]{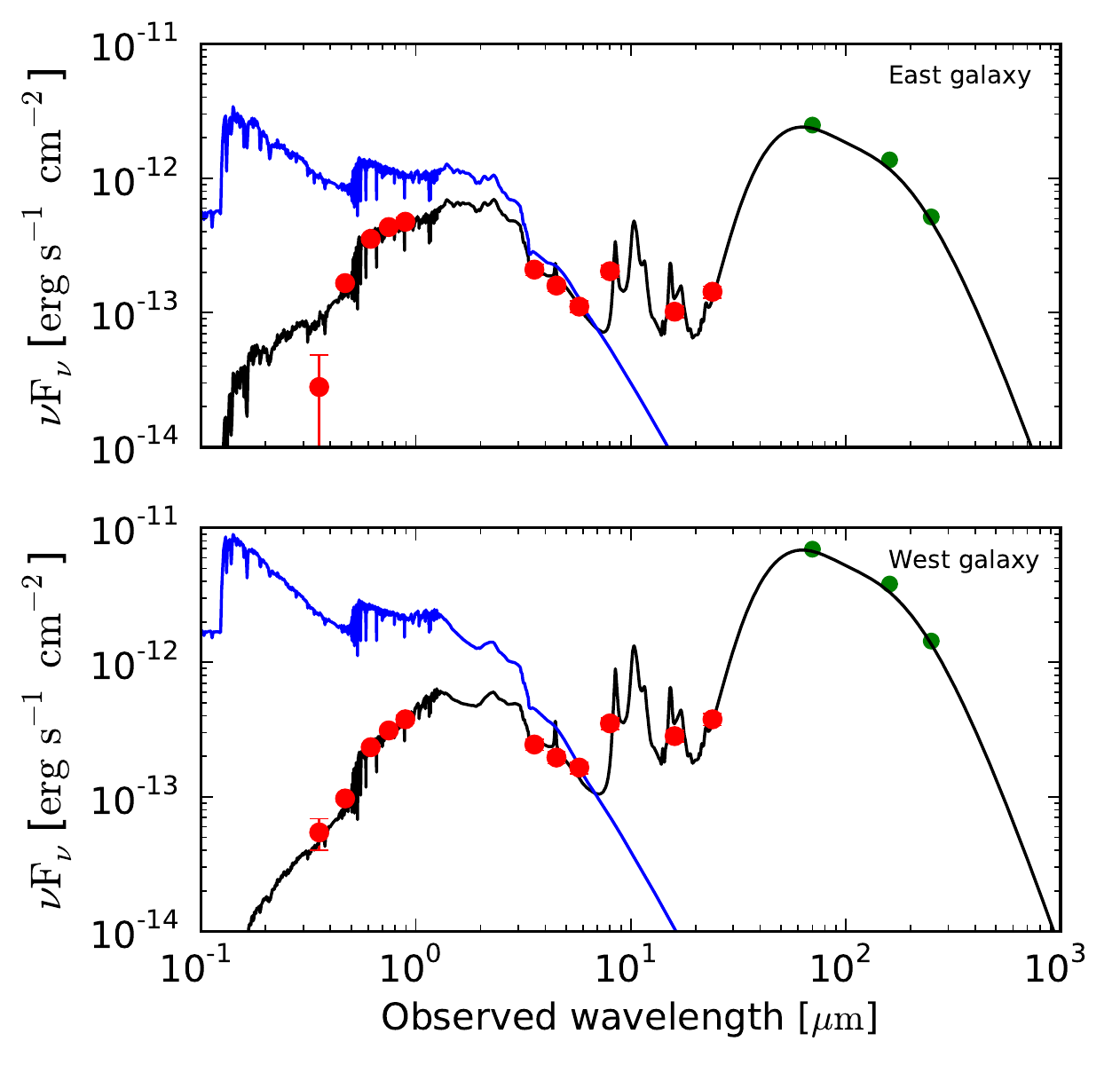}
\caption{MAGPHYS best-fit spectral energy distributions (SEDs) of the interacting pair close 
         (in projection) to 3C~318. Top: the East galaxy. Bottom: the West galaxy. The observed 
         (\textit{Spitzer} and SDSS) photometry is listed in Table~\ref{table:photometry}, and 
         plotted with red circles. The blue and black solid lines, respectively, correspond to 
         the unattenuated and attenuated best-fit SEDs. Green circles represent the photometry at 
         the three shortest \textit{Herschel} bands, estimated using the computed infrared 
         luminosity ratio and the measured integrated \textit{Herschel} flux density (see text).    
        }
\label{figure:magphys}
\end{figure}

The availability of photometric data over a large wavelength range allows us to obtain 
reasonable estimates of the galaxies' physical properties using an SED fitting approach. 
To this end, we use the code MAGPHYS\footnote{Publicly available at 
\url{http://www.iap.fr/magphys/magphys/MAGPHYS.html}}, which follows the approach 
outlined by \citet{daCunha08}. This largely empirical (but physically motivated) code 
interprets the infrared SEDs of galaxies consistently with the emission at shorter 
wavelengths assuming (1) that only starlight heats the dust and (2) that the luminosity 
absorbed by dust is re-emitted in the infrared domain. For details related to the 
adopted stellar populations and dust attenuation we refer the reader to \citet{daCunha08}.   
Using $\chi^2$ minimization, the code constrains several physical properties related to 
the stars and the interstellar medium of the probed galaxies. 
\begin{table}
\renewcommand{\arraystretch}{1.2}
\caption{MAGPHYS Best-fit physical properties.}   
\label{table:results}      
\centering                      
\begin{tabular}{ccccc}      
\hline\hline
            & M$_{\mathrm{stellar}}$ & L$_{\mathrm{IR}}$      & M$_{\mathrm{dust}}$  & $\chi^2$ \\
            & [M$_{\odot}$]          & [L$_{\odot}$]          & [M$_{\odot}$]        & \\
\hline\hline
East galaxy & 5.0 $\times$ 10$^{10}$ & 3.9 $\times$ 10$^{11}$ & 1.6 $\times$ 10$^{8}$ & 2.7 \\
West galaxy & 8.0 $\times$ 10$^{10}$ & 1.1 $\times$ 10$^{12}$ & 4.6 $\times$ 10$^{8}$ & 1.3 \\
\hline          
\end{tabular}
\end{table}
Figure~\ref{figure:magphys} shows the best-fit SEDs for the two interacting galaxies as 
determined by MAGPHYS using the SDSS and \textit{Spitzer} photometry listed in 
Table~\ref{table:photometry}, and the redshift of the interacting pair determined from 
the optical spectra. The fits are remarkably good despite the lack of constraints in the 
FIR domain, and both galaxies are best fitted with dusty starburst templates. 
Both galaxies are characterized by a considerable amount of dust attenuation and have their 
IRAC 8.0~$\mu$m and IRS 16~$\mu$m fluxes dominated by emission from polycyclic aromatic 
hydrocarbon features. As a result from the low signal-to-noise \textit{u}-band photometry 
for the East galaxy, its best-fit is slightly worse than that of the West galaxy. No 
differences in the derived paramaters were found when performing the fit without taking this 
data point into account. The results, listed in Table~\ref{table:results}, suggest that 
relative to the East galaxy, the West galaxy has a factor of 1.6 higher stellar mass 
(M$_{\mathrm{stellar}}$), and a factor of 2.8 higher infrared luminosity (L$_{\mathrm{IR}}$). 
We adopt the L$_{\mathrm{IR}}$ ratio between the two galaxies, and use it to estimate the 
contribution of each individual galaxy to the integrated flux density measured in the three 
shortest \textit{Herschel} bands (see Table~\ref{table:photometry}). We perform additional 
fits using these estimated flux densities in addition to the SDSS and \textit{Spitzer} flux 
densities, and find similar physical properties (within 10\%) confirming that the fits obtained 
without the \textit{Herschel} constraints are remarkably robust. For completeness, we overplot 
the estimated \textit{Herschel} photometry in Fig.~\ref{figure:magphys} (green points). The 
morphology and size of the interacting pair of galaxies are comparable to those of the well-known 
Antennae \citep[NGC~4038/NGC~4039, e.g.][]{Whitmore&Schweizer95}, which have a total FIR luminosity 
of L$_{\mathrm{FIR}}$ $\sim$ 1.3 $\times$ 10$^{11}$ L$_{\odot}$ \citep{Klaas10}. The total infrared 
luminosity of the pair is in line with \textit{Herschel}-studied ULIRGs at 0.2~$<$~$z$~$<$0.9 \citep{Magdis14}. 

\citet{Willott00} presented the hitherto obtained radio to X-ray SED of 3C~318 and determined 
that its infrared luminosity, L$_{\mathrm{FIR}}$, greatly exceeds 10$^{13}$ L$_{\odot}$. 
Correcting for the non-thermal core emission, \citet{Willott07} modified it to a value of  
L$_{\mathrm{FIR}}$ $\sim$ 10$^{13}$ L$_{\odot}$, placing 3C~318 at the border between what 
is known in the literature as ultra-luminous (ULIRG) and hyper-luminous (HyLIRG) infrared 
galaxies. Using better quality \textit{Spitzer} and \textit{Herschel} data, P15 fitted the 
infrared SED of 3C~318 obtaining L$_{\mathrm{SF}}$ = 3.4$\substack{+ 0.4 \\ - 0.3}$ $\times$ 
10$^{12}$ L$_{\odot}$, firmly establishing the ULIRG nature of 3C~318. 

Here we revisit the infrared SED of 3C~318 using the just obtained SPIRE 350 and 500~$\mu$m 
photometry. To isolate the signal from 3C~318, we use the best-fits for the pair galaxies 
determined by MAGPHYS, extrapolate the flux densities at these two bands, and finally subtract 
them from the integrated values given in Table~\ref{table:photometry}. Following P15, we 
fit 3C~318's infrared SED using the sum of three components accounting for the emission from 
AGN-heated hot dust close to sublimation temperature (represented by a blackbody with a fixed 
temperature of 1300~K), the AGN-heated warm dust (torus) emission \citep[represented by a model 
from the library of][]{Hoenig&Kishimoto10}, and the SF-heated dust on the scale of 
the host galaxy (represented by a modified blackbody with a fixed emissivity index). 

\begin{figure}
\centering
\includegraphics[width=\hsize]{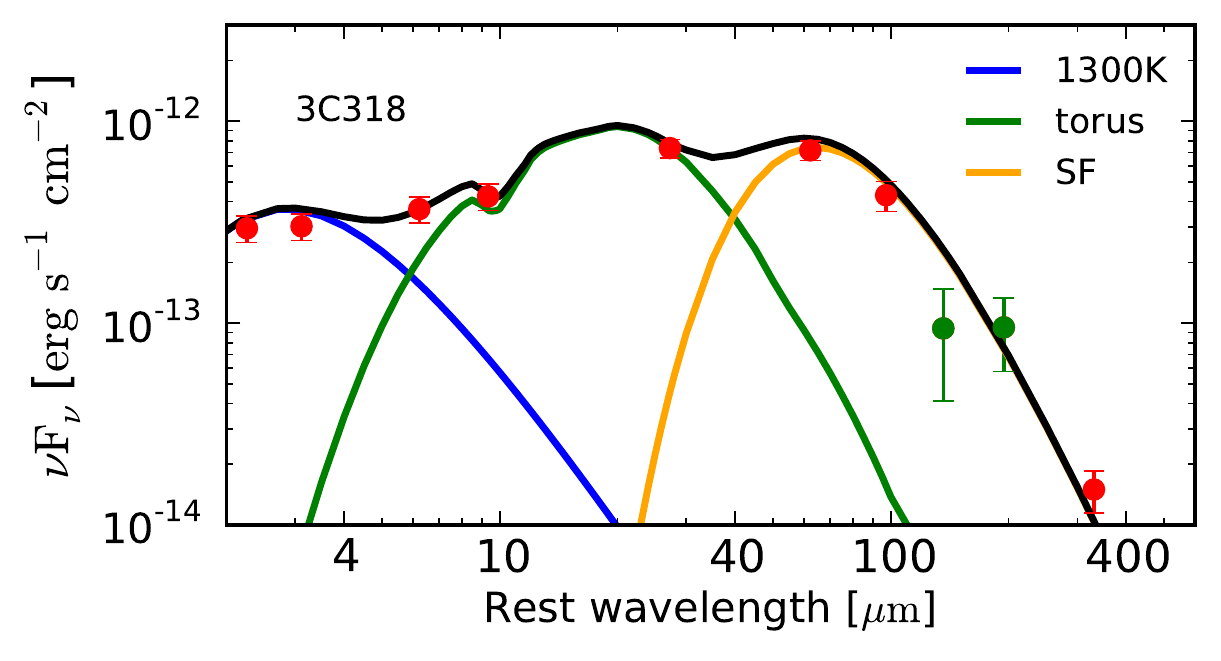}
\caption{Best-fit infrared spectral energy distribution (SED) of the quasar 3C~318. Red 
         circles represent the observed photometry, and green circles the estimated photometry 
         (see text). The total SED (black) is given as the sum of AGN-heated hot dust close to 
         sublimation temperature (blue), of torus emission \citep[green, from the library of][]
         {Hoenig&Kishimoto10}, and of star-formation-heated dust (yellow).         
         }
\label{figure:3c318sed}
\end{figure}

The best-fit SED, presented in Fig.~\ref{figure:3c318sed}, represents the photometric data points 
well, except for the SPIRE 350~$\mu$m point. Integrating under the AGN and SF related 
components gives L$_{\mathrm{AGN}}$ = 7.6 $\times$ 10$^{12}$ L$_{\odot}$ and L$_{\mathrm{SF}}$ = 3.4 $\times$ 
10$^{12}$ L$_{\odot}$, comparable to what was found for other SPIRE detected sources from the same 3CR catalogue (P15).  
The L$_{\mathrm{SF}}$ value obtained here is a factor of three lower than that determined by 
\citet{Willott07}. Consequently, the gas-depletion time scale (or SF efficiency) 
of 3C~318 is a factor of three longer than that computed by \citet{Heywood13} based on a 
molecular gas mass of $M_{H_2}$ $\sim$ 3.7 ($\pm$0.4) $\times$ 10$^{10}$ ($\alpha_{CO}$/0.8) M$_{\odot}$, 
and equals about 60 Myr. These numbers place 3C~318 well below the L$_{\mathrm{FIR}}$-L'$_{\mathrm{CO}}$ 
relation for high-$z$ submillimetre galaxies reported by \citet{Bothwell13}, but still imply 
strong SF activity in its host galaxy. Note that part of the measured infrared flux 
density in the \textit{Herschel} bands could also be attributed to the close companion south-west 
of 3C~318, noted by \citet{Willott00} and seen in Fig.~\ref{figure:stamps}. Whether this companion 
is at the same redshift as 3C~318, and possibly interacting with it, remains to be seen. Future 
ALMA studies will likely provide an improved view of the dust and molecular gas emission surrounding 
3C~318. 
\section{Conclusion}
\label{section:Conclusion}  
\textit{Herschel} imaging of the field centered on the radio-loud quasar 3C~318 reveals 
that the infrared emission detected with earlier instruments comes from a pair of interacting 
galaxies close (in projection) to 3C~318. Using optical spectroscopy, we determine the redshift 
of the pair to be $z$=0.35. We perform a spectral energy distribution analysis of the pair, 
and constrain its infrared luminosity to L$_{\mathrm{IR}}$ = 1.5 $\times$ 10$^{12}$ L$_{\odot}$, 
comparable to the luminosity of other ultra-luminous infrared galaxies at that redshift. We  
quantify the emission of 3C~318's host galaxy in \textit{Herschel} bands where it is blended 
with the galaxy pair, and find that its luminosity due to SF activity is 
L$_{\mathrm{SF}}$ = 3.4 $\times$ 10$^{12}$ L$_{\odot}$, a factor of three lower than found in 
previous studies. Further ALMA studies are needed to get an improved understanding of the 
dust emission in 3C~318's host. 
\begin{acknowledgements}
   PP acknowledges the Nederlandse Organisatie voor Wetenschappelijk Onderzoek (NWO) for a PhD fellowship.  
   The authors acknowledge Javier Mendez for obtaining the optical spectroscopy, Steve Willner for providing the \textit{Spitzer} photometry, Scott Trager for helping with the redshift determination, and Wouter Karman for useful comments.
   \textit{Herschel} is an ESA space observatory with science instruments provided by European-led Principal Investigator consortia and with important participation from NASA.
   This work is partly based on observations made with the \textit{Spitzer} Space Telescope, which is operated by the Jet Propulsion Laboratory, California Institute of Technology under a contract with NASA. 
   Funding for the Sloan Digital Sky Survey IV has been provided by the Alfred P. Sloan Foundation and the Participating Institutions. SDSS-IV acknowledges support and resources from the Center for High-Performance Computing at the University of Utah. The SDSS web site is www.sdss.org.
   This research made use of APLpy, an open-source plotting package for Python hosted at http://aplpy.github.com.
\end{acknowledgements}
\end{document}